\documentclass[prl,twocolumn,superscriptaddress,notitlepage]{revtex4-1}

\usepackage{graphicx}
\usepackage[usenames,dvipsnames]{color}
\usepackage{amsmath}
\usepackage{amssymb}
\usepackage{soul}
\renewcommand{\a}{\alpha}
\renewcommand{\b}{\beta}
\renewcommand{\d}{\delta}
\newcommand{\e}{\varepsilon}
\newcommand{\f}{\phi}
\newcommand{\g}{\gamma}

\renewcommand{\k}{\kappa}

\renewcommand{\r}{\rho}

\newcommand{\x}{\xi}
\newcommand{\y}{\psi}

\newcommand{\dl}{\partial}

\newcommand{\comments}[1]{}

\newcommand{\tach}[1]{{\color{CornflowerBlue} #1}}

\begin{document}

\title{Chaoticity without thermalisation in disordered lattices}

\author{O.~Tieleman}
\affiliation{Max-Planck-Institut f\"ur Physik komplexer Systeme, N\"othnitzer Stra{\ss}e 38, D-01187 Dresden, Germany.}

\author{Ch.~Skokos}
\affiliation{Physics Department, Aristotle University of Thessaloniki, GR-54124, Thessaloniki, Greece}
\affiliation{Max-Planck-Institut f\"ur Physik komplexer Systeme, N\"othnitzer Stra{\ss}e 38, D-01187 Dresden, Germany.}
\date{\small\it \today}

\author{A.~Lazarides}
\affiliation{Max-Planck-Institut f\"ur Physik komplexer Systeme, N\"othnitzer Stra{\ss}e 38, D-01187 Dresden, Germany.}

\begin{abstract}
We study chaoticity and thermalization in Bose-Einstein condensates in disordered lattices, described by the discrete nonlinear Schr\"odinger equation (DNLS). A symplectic integration method allows us to accurately obtain both the full phase space trajectories and their maximum Lyapunov exponents (mLEs), which characterize their chaoticity. We find that disorder destroys ergodicity by breaking up phase space into subsystems that are effectively disjoint on experimentally relevant timescales, even though energetically, classical localisation cannot occur. This leads us to conclude that the mLE is a very poor ergodicity indicator, since it is not sensitive to the trajectory being confined to a subregion of phase space. The eventual thermalization of a BEC in a disordered lattice cannot be predicted based only on the chaoticity of its phase space trajectory.
\end{abstract}

\pacs{}
\maketitle

\paragraph{Introduction}
In this Letter, we bring together the topics of disorder, nonlinearity, chaos, ergodicity, and Bose-Einstein condensation (BEC) in optical lattices. Inspired by the realisation of disordered optical potentials in experiments with ultracold atomic gases \cite{bil08,roa08}, we explore the question of thermalisation in such systems. Previous theoretical works have, amongst others, addressed the topics of Bose and Anderson glasses \cite{dam03}, Anderson localisation \cite{and58,sch05}, and Lifshits glasses \cite{lug07}. In \cite{lug07}, various regimes of interaction strengths were investigated, and it was found that for sufficiently strong interactions, a disordered BEC is expected. We will focus on this regime, and study the Bose-Hubbard model \cite{fis89,jak98}, which describes a bosonic gas in a lattice, in the mean-field approximation. The resulting model can be obtained by discretising the Gross-Pitaevskii equation, and is also known as the discrete nonlinear Schr\"odinger equation (DNLS) \cite{kev10}. In this Letter, we pose and answer the following question: what is the effect of disorder on thermalisation and ergodicity in the mean-field Bose-Hubbard model / DNLS?

The connection between chaoticity and thermalisation was recently discussed in the disorder-free case \cite{cas09}. Intuitively, one would expect chaotic trajectories to thermalise, since unlike regular ones, they are not confined to the neighborhood of stable periodic orbits. Not being confined, the expectation is that they are able to cover the available phase space, and that the system is therefore well-described by the microcanonical ensemble, since only the energy and particle number are conserved. We show that this expectation is not correct, by explicitly demonstrating the absence of equiprobability of states on the microcanonical shell. Since the energies involved are higher than the disorder potential, classical localisation cannot be responsible for this effect.

The most commonly employed method of chaos detection, which quantifies the sensitive dependence on initial conditions, is the evaluation of the maximum Lyapunov exponent (mLE) \cite{BGGS80a,BGGS80b,ER85,S10}. In \cite{cas09}, a positive mLE, which indicates a chaotic trajectory, was found to predict thermalisation. To test this hypothesis in the disordered case, we investigate whether trajectories classified as chaotic based on the value of their mLE can be confined to subregions of phase space. Based on numerical studies of the disordered DNLS model, we find that generally, the mLE does not have any predictive power concerning spatial confinement of trajectories. Hence, we conclude that the mLE is not sufficient to answer such questions, since it only tells us whether trajectories behave chaotically in a local sense. To the best of our knowledge, no studies addressing this question of confinement and chaoticity in disordered systems have been done hitherto. Confined trajectories are manifestly not ergodic, and therefore do not correspond to thermalising systems. Consequently, we find that in the presence of disorder, the connection between chaoticity and thermalisation is broken. Eventual thermalisation cannot be predicted based on the mLE alone; it has to be supplemented with at least one other quantity, which characterises the degree of confinement of the trajectory.

An important aspect of our system is the interplay between disorder and nonlinearity, which has been studied extensively in recent years \cite{M98,PP08,kop08,fla09,sko09,VKF09,mul09,sko10a,F10, LBKSF10,BLGKSF11,bod11,B11,ver12,mul09},\cite[Sec.~7.4]{CHD12}. Most of these studies were concerned with the evolution of initially localized wave packets. It has been shown that for moderate values of nonlinearity the wave packets' second moment $m_2$ grows subdiffusively in time $t$, as $t^{\alpha}$ with $0 <\alpha <1$ \cite{fla09}. For weak enough nonlinearities, wave packets appear to be frozen for very long time intervals (whose duration increases as nonlinearity decreases) in a manner resembling Anderson localization. For very strong nonlinearities the existence of self-trapping behavior was theoretically predicted \cite{kop08} and numerically verified \cite{fla09,sko09,LBKSF10,BLGKSF11,bod11} for the DNLS model, i.e.~a part of the wave packet remains localized while the rest spreads subdiffusively. In our study, Anderson localisation plays an important role in understanding the physics.

\paragraph{System}
The Hamiltonian for the disordered DNLS is given by
\begin{align}\label{eq:ham}
H = & \, \sum_j \biggl[ -J \Bigl( \y^*_j \y_{j+1} + {\rm h.c.} \Bigr) + \e_j |\y_j|^2 + \frac{\b}{2} |\y_j|^4 \biggr],
\end{align}
with site index $j$, complex wavefunction $\y_j$, hopping strength $J$, on-site disorder $\e_j \in [-W/2, W/2]$, and nonlinear interaction strength $\b$. We apply periodic boundary conditions. The quadratic part of the Hamiltonian can be written in diagonal form in the basis of its {\it normal modes}, which are all exponentially localised in the presence of disorder - this is the phenomenon of Anderson localisation \cite{and58}. The localisation length $\x$ varies within the band, but is bounded: $\x \lesssim 96 (J / W)^2$ \cite{kra93}. The eigenenergies of the quadratic part lie in the interval $[-z J - W/2, z J + W/2]$, yielding a bandwidth of $2 z J + W$, where $z$ is the number of neighbours per site ($2$, in our case). The importance of the interaction term is quantified by the nonlinearity parameter
\begin{align}\label{eq:k}
\k = (\bar{n} \b)/(J + W/2z),
\end{align}
where $\bar{n}$ is the average density: $\bar{n} = (1/L) \sum_j |\y_j|^2$ in a system of $L$ sites. Note that $\k$ reduces to the interaction/kinetic ratio $\bar{n} \b / J$ in the non-disordered case. The coupling between the normal modes due to the interaction term is proportional to the overlap integral $I_{\a\b\g\d} = \sum_j \f_{\a, j} \f_{\b, j} \f_{\g, j} \f_{\d, j}$, where $\f_{\a, j}$ is the value of the $\a$th eigenvector of the noninteracting Hamiltonian at site $j$. In the limit where the normal modes are each localised on a single site and $\f_{\a, j} = \d_{j, j_\a}$, this coupling vanishes. Thus, the system can be effectively non-interacting even if $\k$ is finite, as long as $\x$ is much smaller than the lattice spacing. We measure the degree of localisation by the the participation fraction (PF) $\f$, which is defined as
\begin{align}\label{eq:pf}
\f(\{ \y_j \}) = \frac{1}{L} \frac{\sum_j |\y_j|^2}{\sum_j |\y_j|^4}.
\end{align}
It indicates the fraction of sites occupied in a state, as its extreme values are $\f=1/L$ for single\tach{-}site occupation and $\f=1$ for equipartition.

\paragraph{Methods}
To obtain the evolution of a state under the Hamiltonian \eqref{eq:ham}, we numerically integrate the coupled ODEs obtained from differentiating the Hamiltonian, $i \dot{\y_j} = \dl H / \dl \y_j^*$:
\begin{align}\label{eq:ode}
i \dot{\y_j} = - J \bigl( \y_{j+1} + \y_{j-1} \bigr) + \e_j \y_j + \b |\y_j|^2 \y_j,
\end{align}
by means of a fourth order symplectic integration scheme explicitly developed for Hamiltonian systems that split in exactly three integrable parts \cite{SGBPE13,GESBP13}. Symplectic integrators are numerical techniques especially designed for the integration of Hamiltonian systems, keeping the error of the computed value of the energy bounded irrespectively of the total integration time (see, for example, Chap.~VI of \cite{Hairer_etal_02}, \cite{MQ06,F06} and references therein). Then we calculate the corresponding mLE using the so-called `standard method' \cite{BGGS80b,S10} which is based on the time evolution of small deviations from the studied state. The  variational equations that govern this evolution are integrated by an extension of the symplectic integrator according to the so-called `tangent map method' \cite{sko10,GES12}. As initial condition, we have used the ground state of the disordered potential combined with a harmonic trap: $V_j = \e_j + V_{\rm trap} [(j-L/2)/L]^2$; the density $S$ was normalised to unity, i.e.~$S = \sum_{j}|\psi_j|^2=1$. The trap was switched off before calculating the evolution, leaving only the disordered potential. Our model has two integrals of motion, as it conserves both the energy (\ref{eq:ham}) and the norm $S$.

\begin{figure}[t!]
\begin{center}
\includegraphics[width=.48\textwidth]{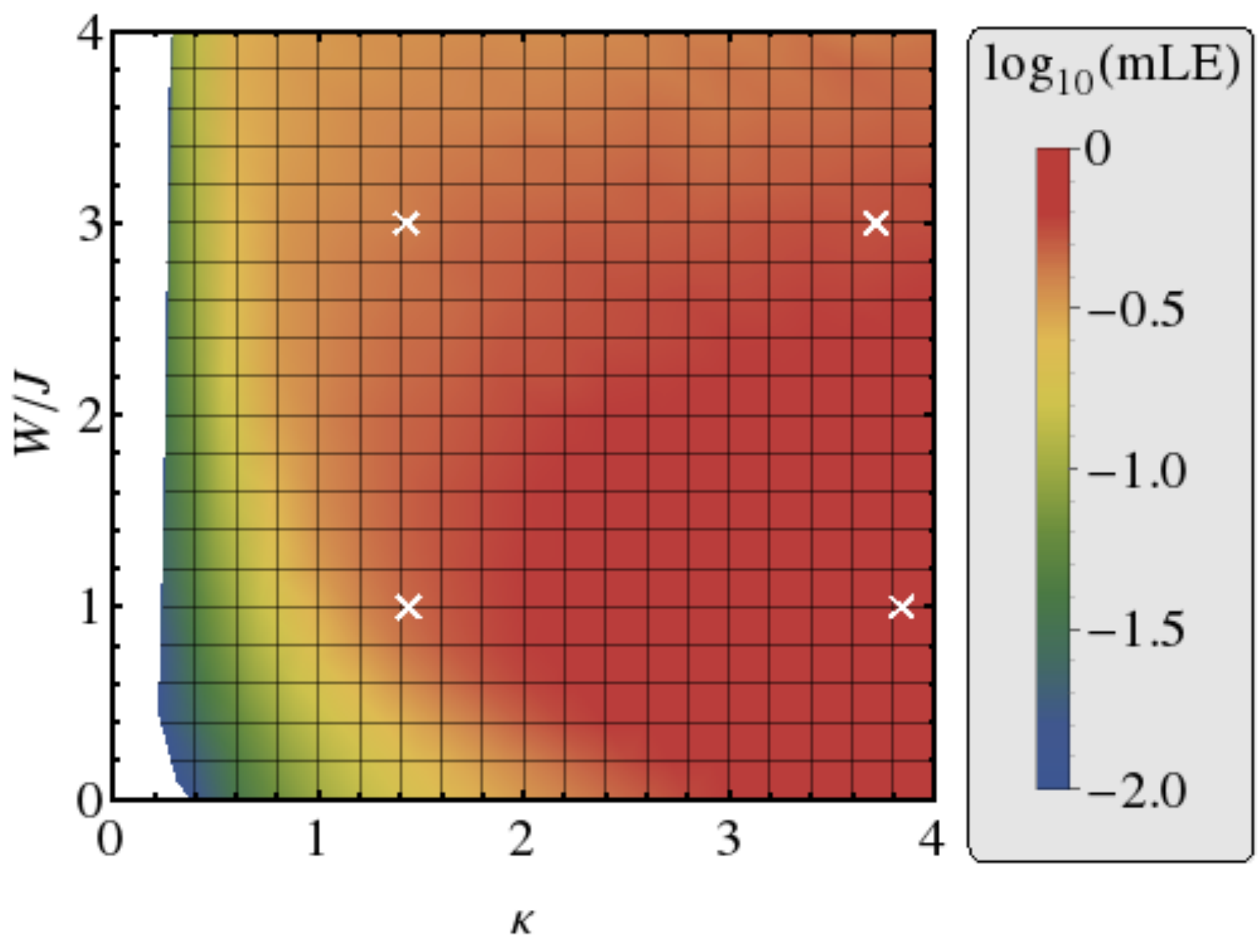}
\end{center}
\caption{The logarithm of the mLE as a function of disorder strength $W/J$ and nonlinearity parameter $\k$. The white crosses indicate the parameter values for the four sets of runs shown in fig.~\ref{fig:scatter}. The white region on the left features very low values for $\log_{10}$ mLE, since the mLE vanishes with $\k$ (see text).}
\label{fig:mLE(W,b)}
\end{figure}

We have generated 100 random disorder realizations, calculated the appropriate initial condition for each realisation and value of $W$, and eventually averaged the mLE outcomes over the different realisations. The results presented here were obtained using a lattice of $L=50$ sites, but the same behaviour was found on lattices up to 500 sites. Each case was integrated up to $t=10^5 \hbar / J$. In all our computations the absolute value of the relative energy and norm error was kept smaller than $10^{-3}$.

Since the tunnelling time is typically a few milliseconds, our integration time of $t=10^5 \hbar / J$ corresponds to hundreds of seconds, while experiments usually do not last longer than a few seconds \cite{gre03}. Our lattice size is also a typical value for experiments. The experimental tunability of the lattice depth via laser power, and atom-atom interaction via Feshbach resonances \cite{chi10}, allow for the exploration of a wide range of disorder strenghts and nonlinearities.

\paragraph{Results}

The disorder-averaged mLE at $t=10^5$ tunnelling time units is shown in log-scale in fig.~\ref{fig:mLE(W,b)}. It increases with the interaction strength $\b$; this may be understood as follows. If the coupling $\b$ vanishes, the Hamiltonian becomes quadratic, and only regular orbits are expected. Switching on interactions induces mixing between the normal modes, and therefore departure from the regular orbits; the higher $\b$, the stronger the mixing and the less regular/more chaotic the trajectories.

The mLE has a non-monotonic behaviour as a function of the disorder strength, i.e.~along vertical lines in fig.~\ref{fig:mLE(W,b)}; the increase for weak disorder is a consequence of our choice of initial condition. For strong enough disorder, any increase leads to a decrease in the mLE, which, as discussed earlier, can be understood as a localisation effect. While all orbits we found at nonzero $\beta$ were chaotic, the range of their final mLE values increases with both W and $\kappa$, indicating the appearance of a mixed phase space. As the localisation of the normal modes increases, both the coupling between them and the value of the overlap integral $I_{\a\b\g\d}$ decrease, leading to a less chaotic system whose regions of regular motion (usually called stability islands) grow in size \footnote{Note that in the extreme case where nonlinearity vanishes the system becomes integrable, chaos is not present and the whole phase space is covered by stability islands (see e.g.~Sec.~1.3 of \cite{LL92}).}. Larger stability islands imply that a larger fraction of chaotic orbits will be influenced by their presence, spending more time close to them, experiencing epochs of less chaotic behaviour, and therefore have a low mLE. The orbits that are in the chaotic sea are not influenced by this effect and therefore are still highly chaotic, leading to a larger spread of mLEs obtained within the sample.
\begin{figure}[t!]
\begin{center}
\includegraphics[width=.48\textwidth]{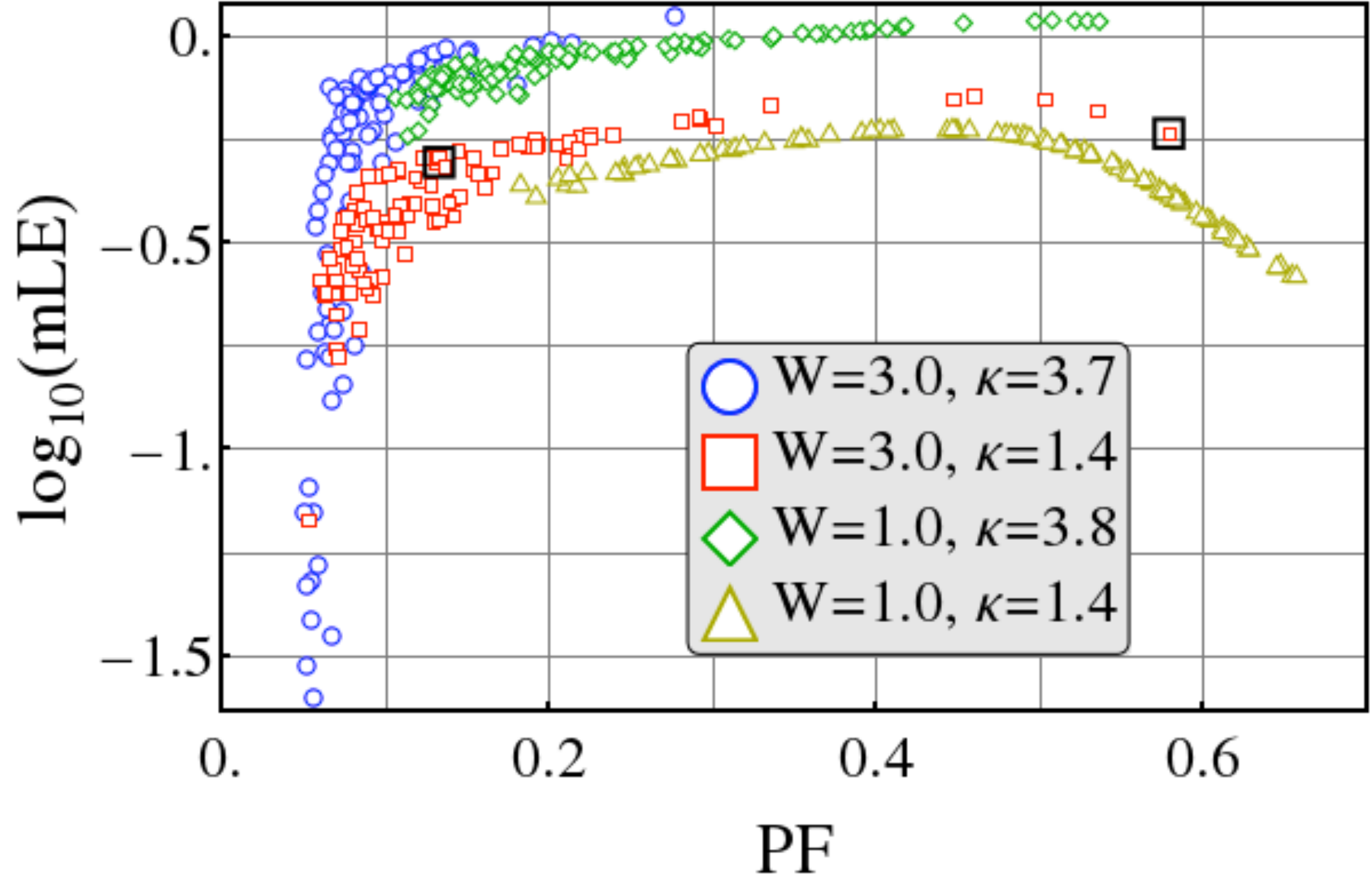}
\end{center}
\caption{The mLE and PF of each of the 100 runs for four combinations $(W, \k)$; $W$ is given in units of $J$. The PF has been averaged over the last 10\% of the run, to smooth out fluctuations (see e.g.~lower panel of fig.~\ref{fig:timeseries}). Two runs are marked by (black) boxes; the time evolution of their PF and density distribution is plotted in fig.~\ref{fig:timeseries}.}
\label{fig:scatter}
\end{figure}

To see how little the mLE tells us about the ergodicity of a trajectory, we now take a closer look at the four $(W, \k)$-combinations that have been marked in fig.~\ref{fig:mLE(W,b)}. In fig.~\ref{fig:scatter}, we plot, for each of these parameter sets, the mLE versus the time-averaged PF $\bar{\f}$ of every run. Three regimes appear in the mLE-PF plot: for $\bar{\f} < 0.1$, a wide range of mLE values occur for very similar $\bar{\f}$; for $0.1 < \bar{\f} < 0.5$, a weakly positive correlation exists between mLE and PF; and for $\bar{\f} > 0.5$, a higher $\bar{\f}$ implies a lower mLE. The results for $\bar{\f} < 0.1$ can be understood in terms of the above-mentioned stability islands. The high-$\bar{\f}$ results present an intriguing topic for follow-up research. Finally, it is noteworthy that the positive correlation between mLE and PF that does exist for intermediate $\bar{\f}$ is very weak.

The most important aspect of fig.~\ref{fig:scatter} for the present discussion is that with the same $W$ and $\k$, significantly different PFs occur for very similar mLEs, as exemplified by the two cases marked in black boxes (both have $W=3$, $\kappa=1.4$). The hypothesised connection between chaoticity and ergodicity would imply that chaotic systems should not be localised, so high mLEs should come with high participation numbers. While such behaviour is found in certain cases for intermediate PF values (between $0.1$ and $0.5$), fig.~\ref{fig:scatter} shows that this is not the case at either low or high values of $\bar{\f}$. We conclude that the mLE has very little predictive power when it comes to localisation.

\begin{figure}[t]
\begin{center}
\includegraphics[width=.48\textwidth]{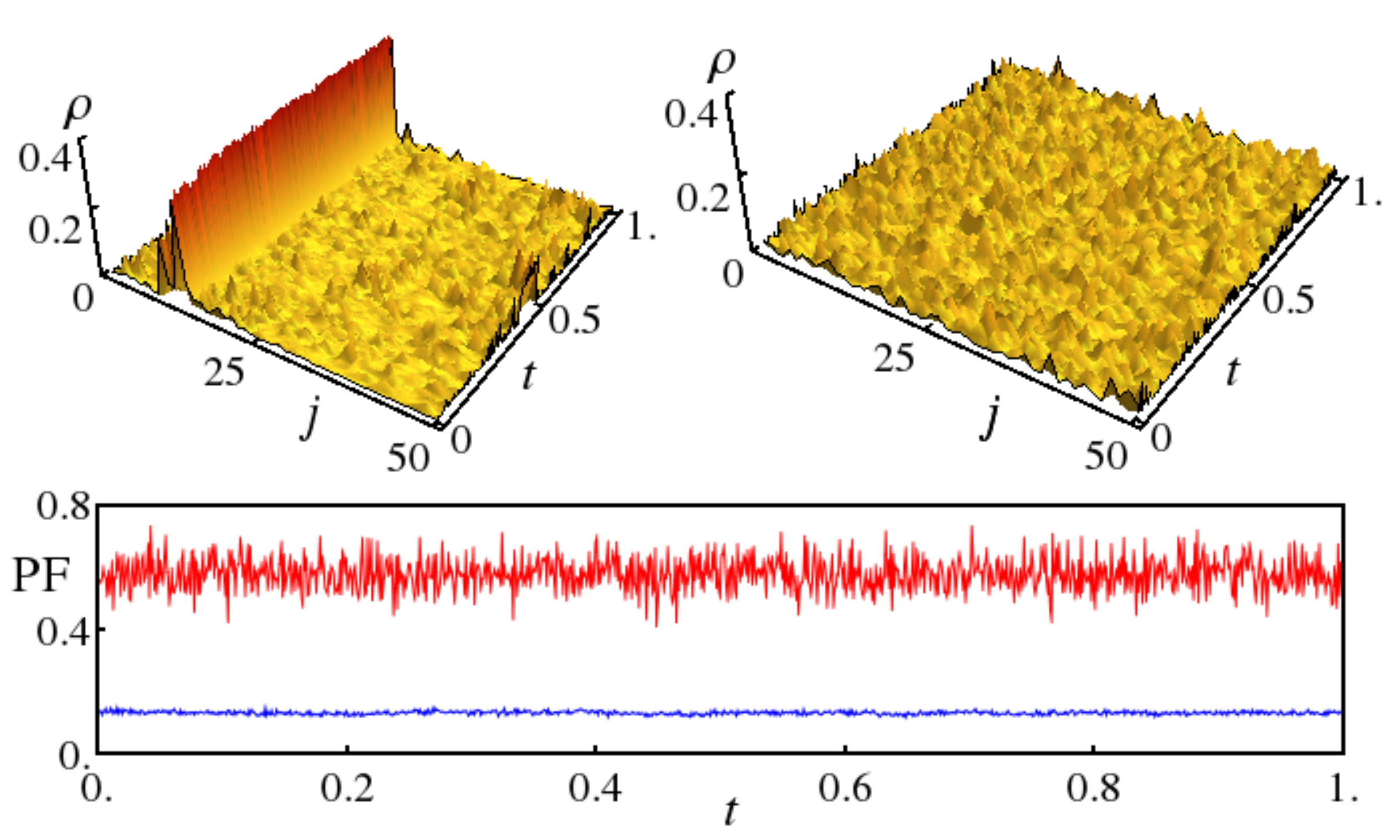}
\end{center}
\caption{The time evolution of the density and PF of two examples, with $W = 3$ and $\k = 1.4$. Time is plotted in units of $10^5 \hbar / J$; $j$ represents the site index, and $\r$ the unit-normalised density. The disorder and interaction strenghts are identical; the only difference is the disorder realisation, and hence the initial condition. The upper right panel shows the density evolution associated with the highest of the two PF values. The energies are $2.42 J$ (upper right) and $0.36 J$ (upper left).}
\label{fig:timeseries}
\end{figure}

Fig.~\ref{fig:timeseries} shows the evolution of the density distribution and PF of the two trajectories marked with black boxes in fig.~\ref{fig:scatter}. The two trajectories have energies of $2.42 J$ (upper left panel) and $0.36 J$ (upper right panel). In one case, a large fraction of the density is stuck on a few sites, and the PF is correspondingly low. States with the same energy but very different density distributions are clearly possible - the peak could occur on different sites with similar potential energy, or be redistributed over sites with higher potential energy. The trajectory fails to reach such states, and is therefore clearly not ergodic, as also indicated by the absence of fluctuations in its PF evolution. This cannot be a classical localisation effect, since the energy of the state is higher than the highest possible on-site potential. We conclude that the chaotic behaviour indicated by the mLE must be confined to a subvolume of the total allowed phase space. The other example does not have any preferred sites, and features much smaller peaks. The evolution of $\f$ for this trajectory shows more fluctuations, indicating more variation in the states it visits, while remaining confined to a rather narrow band.

\paragraph{Discussion/Conclusion}

The observed confinement of trajectories invites more discussion. In two degree of freedom Hamiltonian systems without escapes, the stability islands separate the phase space in non-communicating regions. Consequently isolated chaotic regions (often named chaotic seas) can occur. In higher dimensional systems this separation is not possible, as the stability islands are tori whose dimensionality is not high enough to let them act as barriers between different regions. \cite[Sec.~1.4b]{LL92} So, even in cases where the phase space of high dimensional nonlinear systems is mostly occupied by stability islands, the remaining chaotic regions are interconnected. Consequently, each chaotic orbit can eventually visit every chaotic region in the phase space, although the time for this to happen might become extremely long. This phenomenon is called `Arnold diffusion' (see for example Sec.~1.4 of \cite{LL92} and Sec.~2.11.14 of \cite{Cont2002}), and allows for the theoretical possibility that the confined orbits we find will, at some point much later in time, escape from their region of confinement. However, such considerations are experimentally not relevant, due to the timescales involved. Thus, the introduction of disorder breaks up the phase space up into independent, individually ergodic subsystems, by creating stability islands. Effective ergodicity for physically relevant parameter values is then broken, as can be seen in e.g.~the evolution shown in the upper left panel of fig.~\ref{fig:timeseries}.

The results shown in fig.~\ref{fig:scatter} provide a starting point for a more in-depth discussion of the various types of dynamics that can be expected in BECs in disordered potentials. The different regimes identified in the above discussion of fig.~\ref{fig:scatter} are likely to have interesting physical explanations that shed more light on the behaviour of superfluids in the presence of disorder. We hope to inspire more research in this direction.

In conclusion, we find that Anderson localisation-like effects break the connection between thermalisation and chaoticity in the disordered DNLS / mean-field Bose-Hubbard model. In particular, we have shown that the mLE fails to distinguish between trajectories which explore all energetically allowed areas of phase space and those confined to a subregion. In other words, the mLE \emph{cannot} be used to determine the applicability of the microcanonical ensemble to the disordered DNLS, unlike in the disorder-free case~\cite{cas09}. In order to predict thermalisation, the mLE has to be combined with some other suitably chosen quantity, such as the PF. We conclude that although chaos is a necessary condition for ergodicity, it is not a sufficient one.

\paragraph{Acknowledgments}

O.T. and A.L. thank Evangelos Siminos for inspiring discussions. Ch.S.~was supported by the Research Committee of the Aristotle University of Thessaloniki (Prog.~No 89317).

\bibliography{Biblio}

\end{document}